\documentclass{article}
\usepackage{ijcai24}

\usepackage{times}
\usepackage{soul}
\usepackage{url}
\usepackage[hidelinks]{hyperref}
\usepackage[utf8]{inputenc}
\usepackage[small]{caption}
\usepackage{graphicx}
\usepackage{amsmath}
\usepackage{amsthm}
\usepackage{booktabs}
\usepackage{algorithm}
\usepackage{algorithmic}
\usepackage[switch]{lineno}
\usepackage{amsfonts} 
\usepackage{bbm}
\usepackage{siunitx}
\usepackage{graphicx,subfig}

\title{Bootstrapping OTS-Funcimg Pre-training Model (Botfip) -- A Comprehensive Symbolic Regression Framework}

\author{
Tianhao Chen$^1$
\and
Pengbo Xu$^{1,2}$\and
Haibiao Zheng$^{1}$
\affiliations
$^1$School of Mathematical Sciences, Key Laboratory of MEA(Ministry of Education), Shanghai Key Laboratory of PMMP, East China Normal University, Shanghai 200241, P.R. China. \\
$^2$Shanghai Zhangjiang Institute of Mathematics, Shanghai, 201203, China.\\
\emails
 52205500028@stu.ecnu.edu.cn,
pbxu@math.ecnu.edu.cn,
 hbzheng@math.ecnu.edu.cn
}

\begin{document}

\maketitle

\begin{abstract}
In the field of scientific computing, many problem-solving approaches tend to focus only on the process and final outcome, even in AI for science, there is a lack of deep multimodal information mining behind the data, missing a multimodal framework akin to that in the image-text domain. In this paper, we take Symbolic Regression(SR) as our focal point and, drawing inspiration from the BLIP model in the image-text domain, propose a scientific computing multimodal framework based on Function Images (Funcimg) and Operation Tree Sequence (OTS), named Bootstrapping OTS-Funcimg Pre-training Model (Botfip). In SR experiments, we validate the advantages of Botfip in low-complexity SR problems, showcasing its potential. As a MED framework, Botfip holds promise for future applications in a broader range of scientific computing problems.

\end{abstract}

\section{Introduction}

Throughout the journey of humanity's quest for mathematical understanding and application, the mathematical interpretation of physical phenomena has always been a pursuit of many physicists. The practice of using mathematical formulas to explain the causes of phenomena has been central to physics, engineering, and many other fields. Experts summarize and induce mathematical formulas from phenomena, a method often referred to as empirical formula summarization, and provide theoretical explanations for them.  This process, known in the realm of modern computational research as Symbolic Regression (SR)\cite{billard2002symbolic}, has always held a significant position. The exploration of SR starts from a simple yet profound desire: to reveal the mathematical laws hidden behind numbers and data. Unlike traditional regression methods, SR does not provide a specific model at the outset. Instead, it builds initial expressions by randomly combining basic mathematical elements such as arithmetic operators, analytic functions, constants, and variables. For instance, consider a dataset \(\left\{(x_i,y_i)\right\}_{i=1}^N\). The aim of SR techniques is to discover an optimally concise composite function \(F\) from a specific set of symbols \(\left\{f_j\right\}_{j=1}^M\) (including unary and binary operations) such that it satisfies \(F(x_i) \approx y_i \) for any \(i\in\{1,...,N\}\) as closely as possible\cite{makke2022interpretable}. This is done not just to find a mathematical equation that fits the data and predicts values, but to deeply understand the natural laws and phenomena contained within the data, and to discover the mathematical expressions themselves. 

Classical approaches to solving SR rely on Genetic Programming (GP), an effective evolutionary algorithm inspired by the process of natural selection to iteratively refine candidate solutions\cite{searson2010gptips,haeri2017statistical,uy2011semantically}. GP generates a population of mathematical expressions by randomly combining operators and numerical values. It then creates new expressions through crossover (exchanging parts of expressions) and mutation (randomly altering parts of an expression). Each generation of expressions is evaluated and scored based on their fitness in approximating the data. Up to now, genetic programming remain one of the mainstream research methods in the field of SR, and there have been several representative works in recent years.  For instance, Miles Cranmer, in his seminal work\shortcite{cranmer2023interpretable}, leveraged the Julia library SymbolicRegression.jl to develop PySR, an open-source SR tool characterized by its high customizability and support for distributed computing.  Bogdan Burlacu et al. innovatively utilized C++ to develop the Operon genetic algorithm framework for SR\shortcite{burlacu2020operon}, renowned for its high performance and ability to express diverse algorithmic variants using advanced constructs. From a theoretical standpoint, Chen Qi et al.\shortcite{chen2020rademacher} proposed the use of Rademacher complexity as a training objective in GP. Their empirical studies validated this approach, demonstrating superior generalization capabilities and improved interpretability in the resulting models.
\begin{figure*}[!t]
    \centering
    \includegraphics[width=1\textwidth]{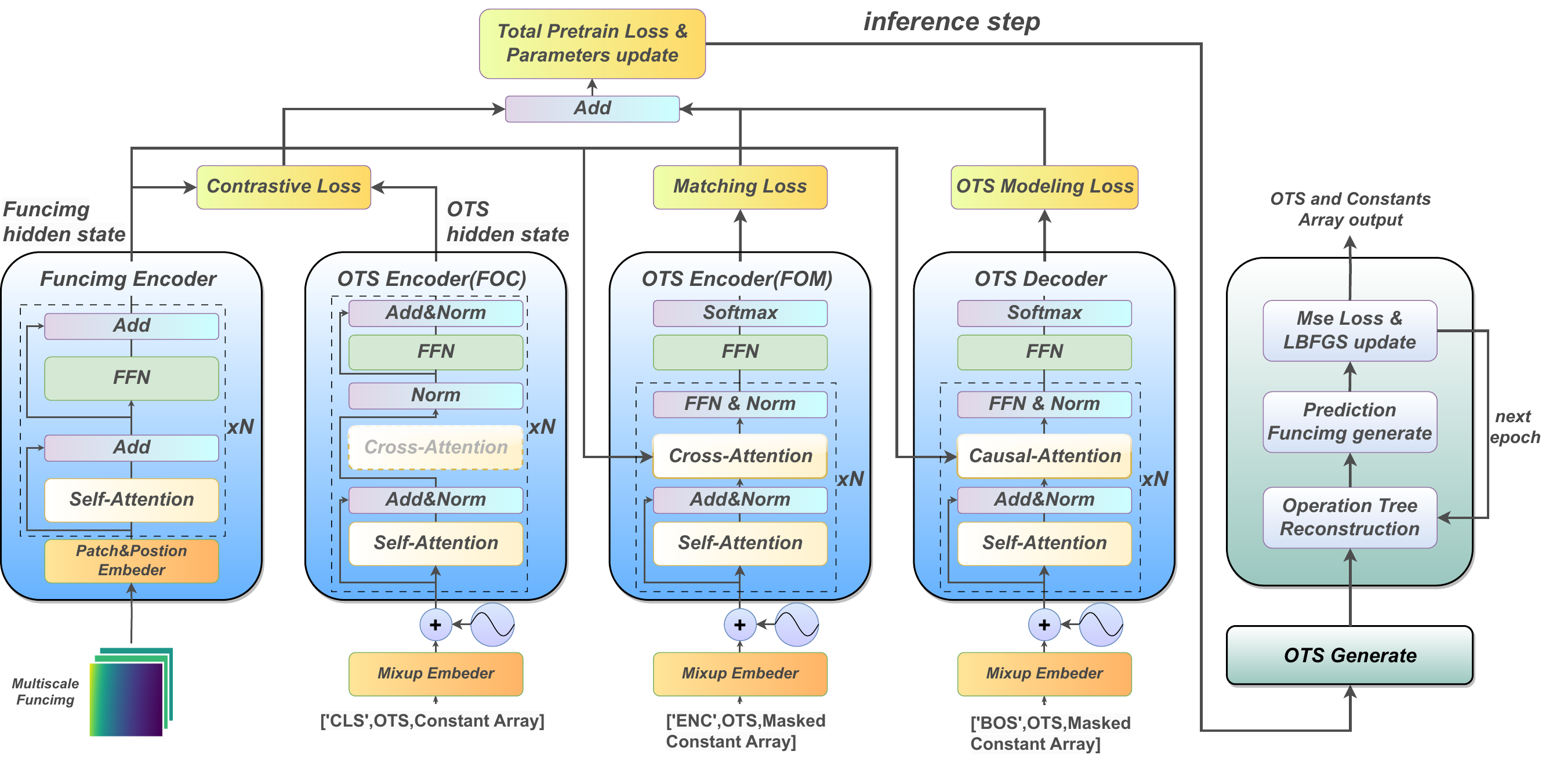}
    \caption{The Botfip framework demonstration, where the left blue part shows the main structure of various components of Botfip (including Funcimg Encoder, OTS Encoder/Decoder) under different pre-training tasks, and also demonstrates the different tasks (FOC, FOM, etc.) in the entire pre-training process. The right green part displays the inference process that model first generate OTS, then output OTS and corresponding constants array obtained through random initialization and iterative optimization using the LBFGS algorithm together.}
    \label{fig:botfip_pretrain_model}
\end{figure*}

In recent years, with the rapid development of artificial intelligence and deep learning, the field of SR has encountered unprecedented opportunities for advancement. Throughout the evolution of artificial intelligence, data-driven methods have increasingly gained prominence. This methodological shift has profoundly influenced SR. In data-driven SR, algorithms no longer rely solely on preset formulas or models, but instead learn and construct mathematical expressions directly from data. This approach shows greater flexibility and effectiveness in solving problems involving nonlinearity, high dimensionality, and complex interactions. Currently, the primary solutions in the field of SR involving artificial intelligence and deep learning are mainly based on reinforcement learning models(RL) and End-to-End models. RL methods enable agents to master optimal decision-making through trial and error within an environment, crafting a training milieu constituted by SR expressions and then incessantly refining their approach to unearth expressions that resonate with the dataset. Contrastingly, End-to-End methods streamline this process by abstracting symbolic expressions into sequences, empowering neural networks to predict these sequences directly from the dataset, thereby deriving symbolic expressions.  Currently, there are several significant work in the field of SR have produced by RL and End-to-End approaches. For instance, Mundhenk Terrell et al.\shortcite{mundhenk2021symbolic} introduced a Reinforcement-Based Grammar-Guided Symbolic Regression (RBG2-SR) method, utilizing context-free grammar a s the action space for RL, thereby constraining the representational space with domain knowledge. Brenden K. Petersen et al. combined recurrent neural networks with risk-seeking policy gradient methods to infer better expressions\shortcite{petersen2019deep}. A representative work in the End-to-End approach is by Kamienny et al.\shortcite{kamienny2022end}, who proposed using Transformer neural networks to directly predict SR expressions from data, achieving significant improvements in inference speed.

While current End-to-End methods have made substantial strides in inference performance, they still face significant limitations when applied to SR problems. Presently, transformer-based End-to-End approaches predominantly focus on leveraging single modality sequential structural features, neglecting to explore other modal features inherent in SR. This has led to a scarcity of derivative applications for End-to-End methods in SR. In recent years, the fields of multimodal learning and contrastive learning have rapidly evolved and playing pivotal roles in areas such as CV and NLP. Notably, in the domain of image-text multimodal learning, the landmark CLIP model \cite{radford2021learning} showcases how encoders for different modalities can encode information, aligning diverse encoded vectors to compute contrastive loss for optimization. This breakthrough has transcended the traditional confines of single-modality feature extraction, elevating the capability to a new level. Furthermore, multimodal learning has significantly expanded the range of tasks in CV and NLP, enabling the realization of zero-shot tasks \cite{cao2023review}, image generation through prompts \cite{lee2023multimodal,khattak2023maple}, among others. A particularly representative task in this context is the Bootstrapping Language-Image Pre-training Model (BLIP) \cite{li2022blip}. Building upon the CLIP contrastive task, BLIP introduces a novel Multimodal Encoder Decoder (MED) architecture. In addition to the Image-Text Contrastive Loss, the pre-training phase of BLIP incorporates Image-Text Matching Loss and Language Modeling Loss to further enhance the quality of pre-training. Multimodal learning not only augments feature extraction capabilities beyond what is achievable in single modalities but also broadens the spectrum of task types. Indeed, one can observe analogous multimodal expressions within SR, revealing the untapped potential of applying these advanced methodologies to this domain.

\textbf{Contributions} In this paper, we introduce the \textbf{Bootstrapping OTS-Funcimg Pre-training Model (Botfip)}, a framework based on function expression sequences and function image multimodality for SR, function identification, and scientific computation, building upon the Blip model. We observe the presence of multimodal expressions in SR, specifically in the relationships between operation trees, function expression sequences and function images. Previous works in the field, particularly End-to-End models like those in \cite{kamienny2022end}, have processed operation trees into sequences for training, similar to Sequence Modeling Tasks in NLP. Yet, these models have not fully harnessed function information, including function images (Funcimg), which are rich in detail and insights. Unlike traditional SR approaches that use datasets of point sets  \(\left\{(x_i,y_i)\right\}_{i=1}^N\), we utilize sophisticated image encoders (e.g., Vision Transformer (ViT), CNNs) for extracting features from function images. This approach enables the extraction of comprehensive global function features, contrasting the local features highlighted by point sets. As in \cite{biggio2021neural}, we separate the \textbf{Operation Tree Sequence (OTS)} from constant vectors, merging them in a sequence encoder before inputting them into a BERT network \cite{devlin2018bert}. Our code implementation and data generation encompass a operation tree random generation system that supports both symbolic (via Sympy) and numerical computations (compatible with numpy and pytorch). This system allows for the creation of custom operation tree generation rules, including operator computations, and the ability to randomly generate operation trees with a predefined number of nodes using Breadth-First Search (BFS) for encoding. It also reconstructs operation trees from OTS sequences and constant vectors. Additionally, we have formulated a universal dataset format for Funcimg-OTS pairs, enabling the generation of datasets of any desired size based on configuration requirements. Unlike the labor-intensive data organization and cleansing in the image-text field, generating large-scale datasets for SR is remarkably straightforward, which inherently positions the Botfip framework as a potential powerhouse for large model applications.

\section{Operation Tree Formulation and Dataset Generation}
\label{sec:Operation Tree Formulation}
\begin{figure}[!t]
    \centering
    \includegraphics[width=1.2\columnwidth]{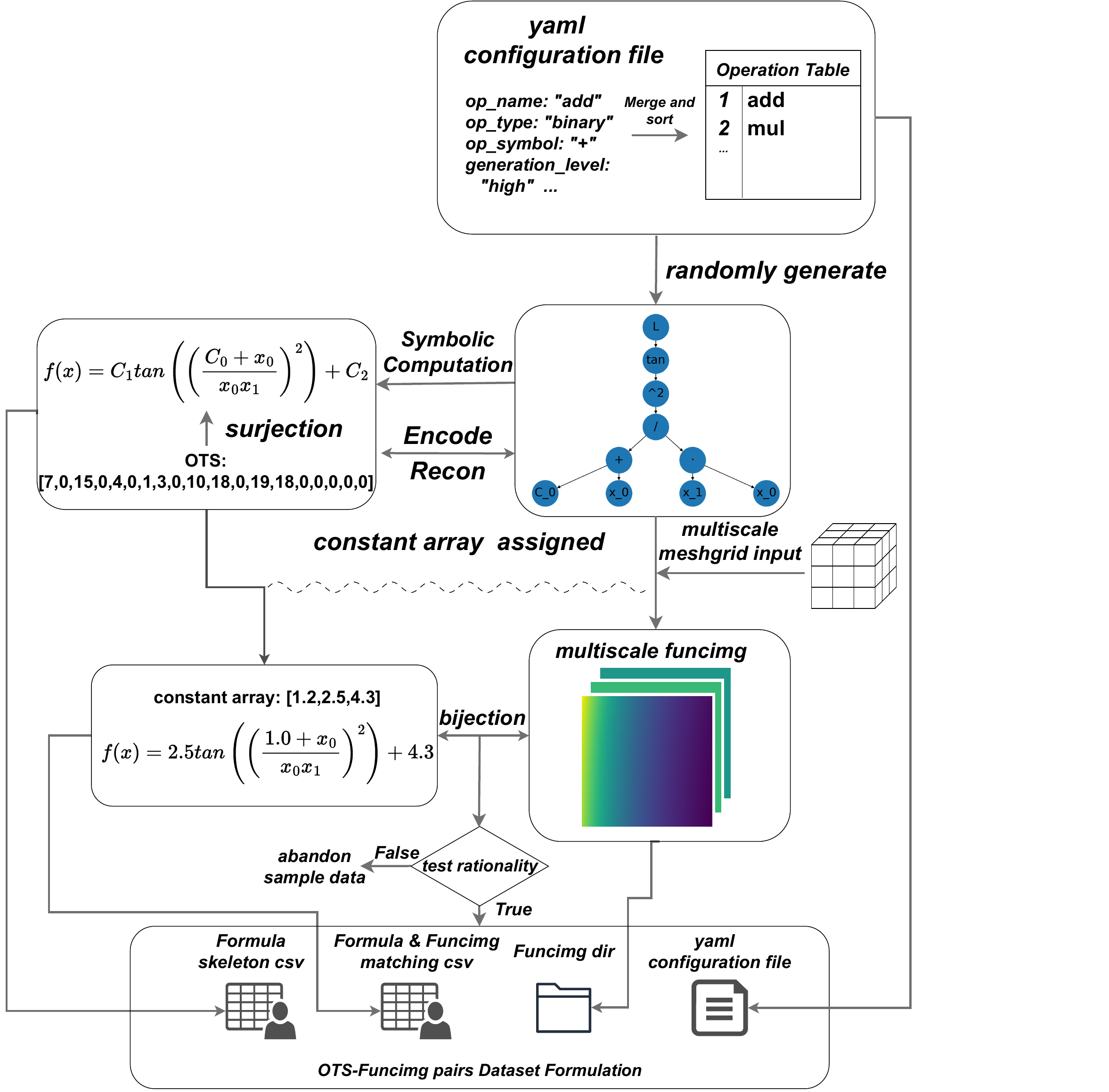}
    \caption{The architecture of the operation tree generation system, the process of generating Funcimg-OTS pairs, and the dataset formulation procedure.
}
    \label{fig:dataset_gen}
\end{figure}

In this section, we introduce the operation tree generation system developed in this study, encompassing various components such as the BFS encoding rules for operation trees, the stochastic rules for generating operation trees, the reconstruction of operation trees from OTS-Constant array, the integration of symbolic and numerical computations, the generation of function images (Funcimgs), and the creation of multi-scale Funcimg-OTS matched pair datasets. Figure \ref{fig:dataset_gen} illustrates the comprehensive process, from the generation of operation trees to the dataset generation through random operation tree sampling.

\subsection{Operation Tree Generation and Reconstruction}
This section primarily describes the operation tree generation system implemented in Botfip. Similar to the methods discussed in \cite{lample2019deep} and \cite{kamienny2022end}, our framework relies on a operation tree system that initially generates random tree structures. Following this, based on a vocab of symbols $\mathcal{V}$, which is isomorphic to the set $ \{0,...,N\}$ and $N$ is the length of vocab $\mathcal{V}$. Each unary and binary node of the operation tree is assigned a symbol. This process culminates in the creation of a corresponding composite function \(f:\mathbb{R}^n \to \mathbb{R}\). However, the distinct aspect of our work lies in the specific approach to operation tree generation and function composition, which diverges from the methods detailed in the referenced works such as:

\begin{itemize}
    \item To enhance the rationality of the formula set generated by our random operation tree system, we have implemented specific constraints on the structure and symbol assignments of the operation trees. During the structure generation phase, for example, we define the maximum permissible length for consecutive appearances of unary operators. Additionally, some symbols are prohibited from appearing simultaneously in adjacent nodes, while others are limited in their frequency of occurrence.  These constraints ensure a balanced and realistic distribution of symbols across the generated operation trees, enhancing the diversity and practicality of the symbolic expressions for regression analysis.  Such an approach ensures the structural integrity and functional diversity of the symbolic expressions, fostering a more robust and versatile dataset for SR tasks.
    \item In our framework, all symbols are treated as operators. Our framework also supports advanced operator symbols for manipulating functions, such as those for differentiation and integration\footnote{However, in our experiments, we exclude differentiation and integration operators as they can lead to overly complex function characteristics.}. 
    \item In the proposed framework, the operation trees generated are ordered following the Breadth-First Search (BFS) approach rather than the Depth-First Search (DFS). In instances where a branch terminates at the same height, an ‘end’ identifier is used to fill the gap in the tree sequence encoding (which will subsequently be replaced by 0). Moreover, our framework implements a distinct separation between OTS and constants arrays. Given the automated recording of the positions and quantities of constants during the tree construction process, combined with the use of BFS encoding, we can effectively decouple OTS and constant values.

\end{itemize}

Based on the aforementioned characteristics, our system allows the construction of a operation tree encoding that is more conducive to language models. For instance, consider the example \(f(x)=\cos(2.4242x)\) from \cite{kamienny2022end}. In our operation tree system, this would be encoded as \([ \text{cos}, *, x, C ]\) along with the constant value \(2.4242\). If we account for the root node symbol being a Linear transformation, the encoding would be \([ L, \text{cos}, *, x, C ]\) with the constant vector \([ 0, 0, 2.4242 ]\). This setup enables us to completely separate OTS from constant values\footnote{It's worth noting that the mapping from OTS to the simplest formula expression is surjective, not bijective, indicating multiple OTS sequences could correspond to the same formula expression.}. Further details about the operation tree construction process and more can be found in Figure \ref{fig:dataset_gen}.

\subsection{OTS-Funcimg Pairs Dataset Generation}

Here we performed the generation of OTS-Funcimg pairs datasets, building upon the operation tree random generation system described earlier. As we established in the previous section, OTS sequences correspond to operation trees. Upon assigning constant vectors, we can obtain forward-computable functions from these trees. Whenever we generate a operation tree with a defined node count to obtain the corresponding function formula \(f(x) : \mathbb{R}^n \rightarrow \mathbb{R}\), we first perform symbolic computation to ensure the rationality of the formula \(f(x)\) produced by the operation tree. Rationality refers to the reasonableness of the formula, such as avoiding infinities and ensuring it is not a mere constant. If the symbolic computation yields a non-erroneous result, we proceed with multi-scale Funcimg generation using the function derived from the operation tree. Depending on the number of variables, we select multiple ranges, such as \([-1,1]^n, [-0.1,0.1]^n\), and \([-10,10]^n\), to generate corresponding meshgrids. The rationale behind using multiple scales is that some functions show little variation at certain scales and their characteristics are better revealed at larger scales. In contrast, others exhibit significant changes at smaller scales, reflecting their functional properties more effectively. Inputting different scale meshgrids into the function yields multi-channel function images at various scales. Prior to further processing these multi-scale funcimgs, they undergo normalization and removal of NaN values.

Following the procedures outlined previously, we successfully obtain the requisite OTS-Funcimg pairs. Instead of generating Funcimgs during model training through operation tree sampling, we pre-generate these images for the pre-training dataset to accelerate the training process. The pre-training dataset comprises four key components: First, it includes a YAML configuration file related to the dataset. Second, there is a Formula Skeleton CSV file, which is compiled from multiple operation tree samplings and serves to record the structural information of the operation tree skeletons. Third, we have a CSV file that matches Formulas with Funcimgs. Lastly, the dataset includes storage files for Funcimgs. The complete process of dataset generation is detailed in Figure \ref{fig:dataset_gen}.

\section{Methods}

In this chapter, we focus on introducing the foundational architecture of the Botfip framework, which encompasses several critical components including the Botfip tokenizer, Mixup Embedder, OTS encoder-decoder, and Funcimg encoder. Additionally, the section delves into the operational processes of both pre-training and fine-tuning tasks. The comprehensive framework and the pre-training workflow can be referenced in Figure \ref{fig:botfip_pretrain_model}.

\subsection{Pretrain and Finetune Tasks}
The Pretrain Tasks primarily comprise Funcimg-OTS Contrastive Training (FOC), Funcimg-OTS Matching Training (FOM) and OTS Modeling Task. FOC focuses on calculating contrastive loss between global states (the first vector in the encoded sequence) produced by Funcimg Encoder and OTS Encoder for their respective inputs, thereby enhancing the similarity in encoded information of corresponding Funcimg, OTS, and Constant arrays. FOM emphasizes determining the match between Funcimg and a given OTS sample, with positive samples (matched) having a target of 1 and negative samples (unmatched) a target of 0; during this, the Constant array is masked. In both FOC and FOM, a queue is used to store encoding vectors of historical training data through rolling updates, providing negative sample data for FOC and FOM. In fact, FOC and FOM are the primary tasks in pre-training. Additional auxiliary tasks can be added as needed, serving as potential fine-tuning tasks later. The OTS Modeling Task involves predicting the next position's token in an OTS sequence input to the OTS Encoder, achieved by comparing with a shifted version of the original sequence using the OTS Decoder combined with the encoded hidden state of Funcimg. 

For the fine-tuning tasks,  we opt to freeze the Encoder and primarily train the Decoder. The training dataset for fine-tuning (Formula Skeleton) is smaller. In fact, the dataset we use in the fine-tuning phase will consist of a set of symbol trees with a specific number of nodes. This is attributed to the direct correlation between the number of nodes in different operation trees and the corresponding length of their OTS. The greater the number of nodes, the more complex the symbolic expression becomes, thereby significantly increasing the complexity of the OTS. In practice, we often assume a probable length for the symbol, and overly complex regression expressions are not applicable. We will also demonstrate in the subsequent fine-tuning experimental results how the model performs on the Funcimg-OTS pairs dataset with operation trees of different node numbers.

\subsection{Model Structure Description}

\textbf{Tokenizer and Mixup Embedder} In the tokenizer, the symbol set vocab \(\mathcal{V}\) is further expanded to include special characters such as [CLS], [ENC], [BOS] (typically added at the start of the sequence), etc. The tokenizer's primary role is to pad the OTS and constants array based on the maximum sequence length \(D_s\) and maximum constant number \(D_c\). It also modifies or inserts special characters in the OTS as required, producing the processed OTS sequence, constant vector, and attention mask \footnote{The BFS encoding process of the operation tree can also be considered part of the tokenizer.}. The Mixup Embedder mainly embeds the processed OTS and constants array together. As the OTS is an integer type while the constants array is real-valued, they are embedded separately and then merged. The OTS embedding follows the word embedding method common in NLP, whereas the constants array passes through a simple feed-forward layer to process the resulting hidden layer vector. Hence, the Mixup Embedder can be viewed as a mapping \(E: \mathcal{V} ^{D_{s}} \times \mathbb{R}^{D_c} \to \mathbb{R}^{d_e(D_s+D_c)}\), where \(\Tilde{\mathcal{V}}=\{0,...,N+\Tilde{N}\}\)  is the extended vocab set, \(\Tilde{N}\) is the number of special token and \(d_e\) is the embedding hidden layer dimension.

\textbf{Funcimg Encoder}  The Funcimg Encoder typically comprises standard image processing networks, such as ViT or CNN. In this paper, we choose ViT as the primary backbone for the Funcimg Encoder. 

\textbf{OTS Encoder and Decoder} The core structure of the OTS Encoder and Decoder is based on the BERT model, operating similarly to that in the BLIP model. During the pretraining phase, we opt for weight tying between the BERT models in the OTS Encoder and Decoder, which helps in reducing the resources required for pretraining. As seen in Figure \ref{fig:botfip_pretrain_model}, there appears to be a structural difference in the OTS Encoder across the FOC and FOM tasks; however, the main body remains unchanged. In the FOC task, the cross-attention module is not activated, whereas in the FOM task, it incorporates cross-attention with the introduction of Funcimg's hidden state vector. Additionally, the FOM task integrates an extra MLP head for classification. Within the OTS Decoder, the attention mechanism in its core transitions from cross-attention to causal attention mechanism \cite{yang2021causal}. Unlike cross-attention that focuses on global sequence attention computation, causal attention is more attuned to the causal ordering of sequences, making it better suited for Sequence Modeling tasks.

\subsection{L-BFGS Update Approach}
In this framework, to ensure that the OTS inference results and the corresponding constants match the Funcimg as closely as possible during the inference phase, we consider initializing the constants array randomly as trainable parameters on top of OTS Modeling. We then reconstruct the operation tree with the current constants array, input the corresponding area meshgrid points into the operation tree to obtain results, and compute the mse loss against the corresponding point values of Funcimg. Following this, the LBFGS optimizer is used to update the parameters. This approach significantly enhances the matching degree of Funcimg, OTS, and Constants Array during the inference process.

\section{Experiment Results}

\subsection{Dataset and Experiment Detail Description}
In this section, we briefly introduce the dataset and some experimental details used in our study. The core model for the OTS encoder/decoder in our research is a 10-layer BERT model, with a hidden layer size of 768. For the Funcimg Encoder, a 10-layer ViT model is employed, also with a hidden layer size of 768. The dataset used in our experiments was generated following the methodology outlined in Section\ref{sec:Operation Tree Formulation} and illustrated in Fig.\ref{fig:dataset_gen}. The hierarchical steps in dataset generation primarily involve determining the range of node numbers for operation trees, the number of tree structure skeletons, the variety of symbols assigned to each skeleton to generate OTS, and finally, sampling a set number of constants arrays to generate Funcimg using the operation trees represented by OTS. Our pre-training dataset was created with node ranges from 5 to 15, comprising 11028 OTS skeletons, and 551400 Funcimg-OTS pairs with sampling range of the operation tree constants array from \(-2\) to \(2\). The validation dataset, with the same node range, included 2941 OTS skeletons and 88230 Funcimg-OTS pairs. Each validation process used 4000 Funcimg-OTS pairs for analysis. During the pre-training process, we employed a warm-up phase along with StepLR for dynamic learning rate adjustment. The initial learning rate during the warm-up was set to \num{1e-6}, with the maximum initial learning rate at \num{1e-4}. The learning rate followed a step schedule with a period of 5 epochs and a decay ratio of 0.9. The total duration of training was extended across 100 epochs. For  validation, assuming that the predicted OTS set and target OTS set in the validation process are \(\Omega = \left\{(p_i,t_i)_{i=1}^N\right\}\), we mainly focused on the OTS generation regularity \(Acc_{r}\), relative sequence Levenshtein similarity \(S_{RL}\) and relative formula-string Levenshtein similarity \(\Tilde{S}_{RL}\) :
\begin{align}
    Acc_{r}\left(\Omega \right) &= \frac{1}{N} \sum_{i=1}^N \mathbbm{1}\left( A_i \right),\\
    S_{RL}\left(\Omega\right)&=\frac{1}{N} \sum_{i=1}^N \frac{l(t_i) - D_{RL}\left(p_i,t_i\right)}{l(t_i)}\\
    \Tilde{S}_{RL}\left(\Omega\right)&=\frac{1}{N} \sum_{i=1}^N  \frac{\mathbbm{1}\left( A_i \right)\left(l(\Tilde{t}_i) - D_{RL}\left(\Tilde{p}_i,\Tilde{t}_i\right)\right)}{l(\Tilde{t}_i)}
\end{align}
where \(A_i\) represents an event that \(p_i \) can be reconstructed as an operation tree, \(D_{RL}\) is the Levenshtein distance \cite{yujian2007normalized}, \(l(t_i)\) represents the length of OTS \(t_j\) and \(\Tilde{p}_i\) and \(\Tilde{t}_i\) are represented as the formula string expressions obtained through symbolic computation after reverting the OTS to the operation tree. Among these metrics, \(Acc_{r}\) represents the extent to which the generated OTS set can be restored to a operation tree structure, serving as a measure of the sequence's regularity. A lower \(Acc_{r}\) indicates a tendency of the model to generate sequences that are irregular (i.e., cannot be restored to a operation tree), which is undesirable. \(Acc_{r}\) illustrates the lower bound of the model's predictions (i.e., output regularity), while \(S_{RL}\) demonstrates the model's capability to identify funcimg and \(\Tilde{S}_{RL}\) not only combines \(Acc_{r}\) and \(S_{RL}\), but also indicates the differences between the target and predicted symbolic expressions after reverting to and computing with the operation tree. From the perspective of SR, this metric is the most rational. For each validation, we sample 4000 validation set samples for testing.

\subsection{OTS Modeling Performance}
\begin{figure}[!t]
    \centering
    \subfloat[]{\label{fig:validation_acc_node_num}\includegraphics[width=0.8\columnwidth]{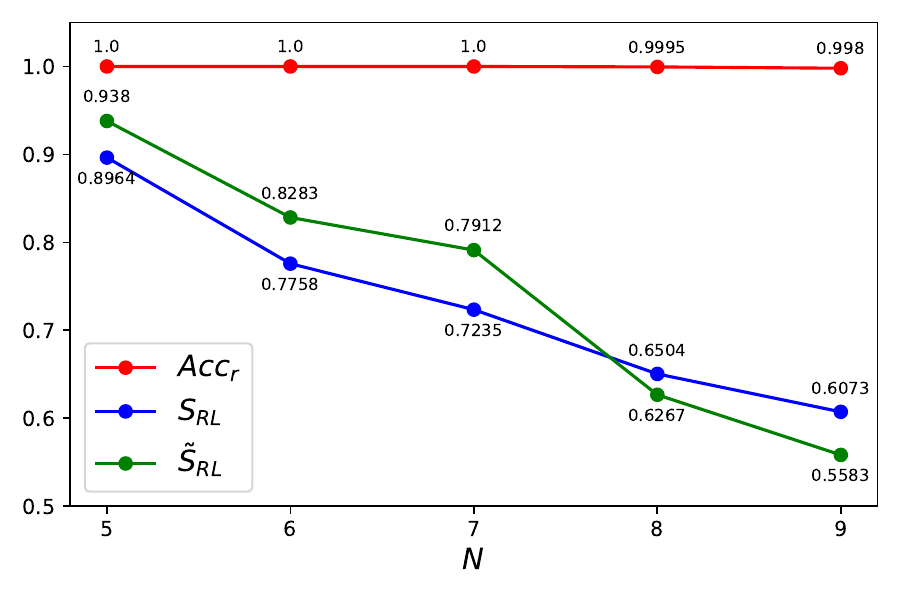}}\\
        \subfloat[]{\label{fig:validation_acc_noise}    \includegraphics[width=0.8\columnwidth]{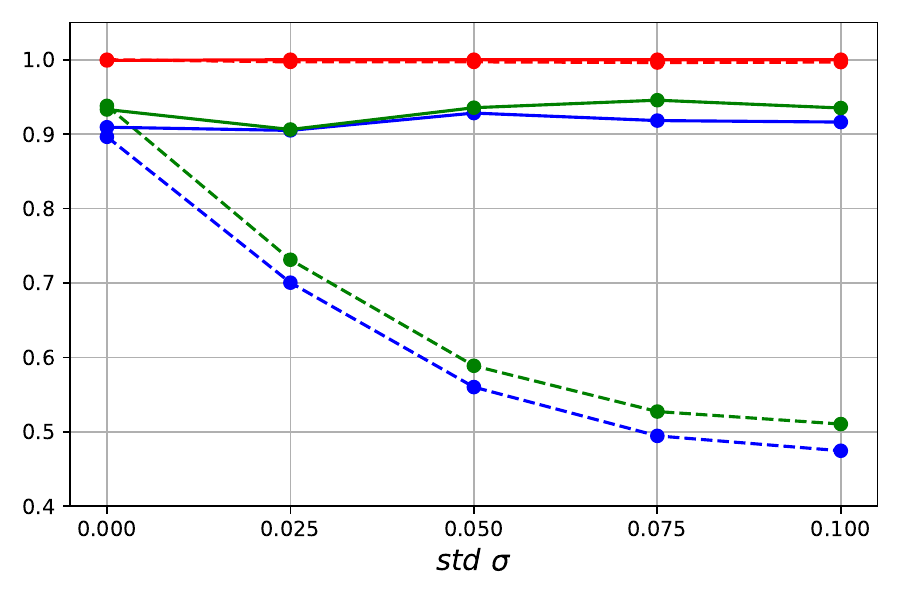}}
    \caption{ Validation results visualization after  OTS fine-tune modeling task. Specifically, Fig.  \ref{fig:validation_acc_node_num} presents the numerical curve of the model's validation loss across datasets composed of operation trees with varying numbers of nodes. Fig.\ref{fig:validation_acc_noise} shows the curve of validation results on a five nodes dataset with input noise. The dashed and solid lines respectively represent the trained model with and without input noise during the pre-training and fine-tuning phases.
}
    \label{fig:validation_acc}
\end{figure}

 we first analyze the OTS inference results of the model on the validation dataset after undergoing pre-training and fine-tuning. The initial test was conducted on the variation of Accuracy for the fine-tuned model under validation datasets composed of different numbers of nodes. As illustrated in Fig.\ref{fig:validation_acc_node_num}, the model's accuracy changes with the number of nodes. The figure demonstrates that when the generated dataset's operation tree has fewer nodes, resulting in shorter OTS sequences, the overall accuracy is higher. The results of \(Acc_{r}\) indicate that our model maintains excellent regularity even with an increasing number of nodes, ensuring reliable model outputs. The results for \(S_{RL}\) and \(\Tilde{S}_{RL}\) reveal that the overall performance of the model's OTS sequence outputs decreases with an increase in the number of operation tree nodes, as the length of the OTS sequence directly correlates with the number of nodes, and longer sequences lead to reduced prediction accuracy. Additionally, longer OTS sequences require more data for alignment and learning, significantly increasing the need for multimodal data. Moreover, \(\Tilde{S}_{RL}\), generally speaking, does not differ much from \(S_{RL}\) in values, but due to the longer formula string sequences, \(\Tilde{S}_{RL}\) is more sensitive to prediction accuracy. Fig.\ref{fig:validation_acc_noise} reflects the model's robustness. It shows that, during the pre-training and fine-tuning phases, if no white noise is added to Funcimg, the model's performance decreases with increasing input noise intensity (i.e., increasing standard deviation), as depicted by the dashed line. In contrast, if a certain level of white noise is artificially introduced during these phases, overall performance and robustness significantly improve, demonstrating the Botfip model's inherent strong generalization and robustness.

\subsection{ L-BFGS Update Performance}
\begin{figure}[!t]
    \centering
    \includegraphics[width=0.8\columnwidth]{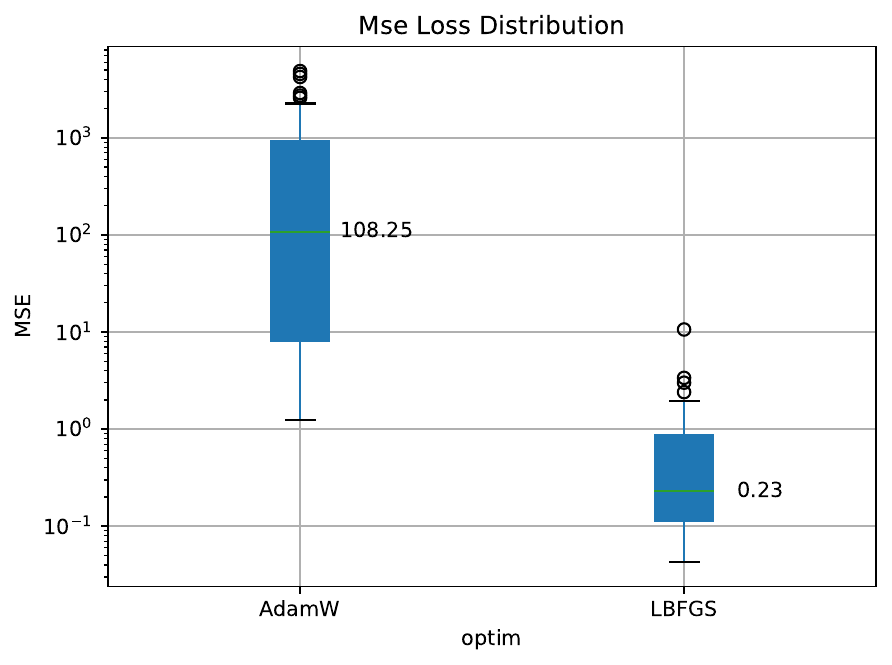}
    \caption{Comparison of Constants Array Update Results between AdamW and LBFGS Optimizers.}
        \label{fig:const_mse_loss_distribution}
\end{figure}
Here we demonstrate the update performance of LBFGS optimizer for constants arrays in multiple Operation Symbolic Trees corresponding to OTS, and compare it with the AdamW optimizer. A validation dataset with node count of 5 is used for testing, setting the batch size to 20 (i.e., generating 20 groups of operation trees at a time)\footnote{This also indicates that our model is capable of batch SR tasks, marking a significant distinction from the traditional GP.}, with a total of 50 sets of test data sampled. Each set is configured with a learning rate of 1, and the stopping threshold(Difference between two training losses) is set at \num{1e-5}. Fig.\ref{fig:const_mse_loss_distribution} presents the testing results. Both types of optimizers reached convergence within just a few epochs for almost all test samples. The distinction lies in the final reconstructed mse loss, where the median loss with the LBFGS optimizer was 0.23, compared to 108.23 with the AdamW optimizer, a difference of three orders of magnitude. This indicates that using the LBFGS optimizer for updating iterations of the constants array in operation trees is highly efficient and precise in SR tasks.

\subsection{Comparison with the Benchmark Method on In-domain Performance}
In this section, we compare some benchmark methods in SR, such as the gplearn framework, PYSR\cite{cranmer2023interpretable}, PSTree\cite{zhang2022ps}, DSR\cite{landajuela2022unified}, and AI-Feynman (AIF)\cite{udrescu2020ai}, focusing on in-domain comparisons. The validation data consists of 100 randomly sampled points from the Funcimg-OTS dataset with node numbers of 5, 7, and 9, representing varying levels of difficulty in SR tasks. Additionally, it is noteworthy that GP frameworks like PYSR often provide multiple potential solutions within a population, where the solution with the minimal fitness tends to be overly complex, showing an overfitting behavior. Therefore, for PYSR and other GP frameworks that offer multiple solutions, we opt for the result that best aligns with the target formula, rather than the one with the lowest fitness.

\begin{table}
    \centering
    \begin{tabular}{lrrr}
        \toprule
         node num &  5& 7 &  9   \\
        \midrule
        gplearn     & 0.535       & 0.663  & 0.754 \\
         PYSR & 0.226          & 0.321  &    0.374 \\
        PSTree & 0.382          & 0.417  &  0.503    \\
        DSR & 0.172         & 0.344 &  0.419  \\
        AIF & 0.262 & 0.423& 0.477\\
        Botfip(Ours) & 0.112& 0.307&0.471\\
        \bottomrule
    \end{tabular}
    \caption{Comparison of Mse  Results on Test Data Across Multiple SR Frameworks}
    \label{tab:validation compared results}
\end{table}
Table \ref{tab:validation compared results} presents the test results of multiple SR frameworks. As can be observed from the table, our Botfip framework performs exceptionally well on Funcimg test sets with a smaller number of nodes, followed by the DSR and PYSR frameworks. However, on test sets with a larger number of nodes (i.e., more complex formula sets), our model's performance is moderate, falling behind DSR and PYSR.  Furthermore, our type of End-to-End framework boasts significant efficiency advantages in the inference process (under GPU computation, one batch of inference process takes almost only 0.1s).

\subsection{Discussion and The Problem of Extrapolation }

Combining the above experimental results, let us summarize the advantages and disadvantages of Botfip in SR tasks. First, GP like those in the gplearn framework tend to generate more complex operation tree structures (corresponding to the length of OTS sequences). For instance, for the target formula \(f(x) = -1.15 \frac{x_0}{x_1} -1.13\), its OTS sequence is \([6,0,4,0,19,18,0,0,0]\). Botfip's corresponding OTS sequence for \(f\) is \([3,0,6,18,0,19,0,0,0]\), which, despite deviating from the original OTS, merely involves the exchange of two adjacent unary and binary nodes and an operation. The final result under BFGS parameter iteration is \(f(x) = -1.15 \frac{x_0}{x_1} -1.1299\). However, the result obtained by gplearn is \(sub(div(inv(-0.875), div(X_1, X_0)), inv(0.895))\), which simplifies to \(f(x) = -1.14 \frac{x_0}{x_1} -1.117\), closely resembling the target formula, but the tree structure generated by gplearn is much more complex, far exceeding our framework. This phenomenon occurs because the Transformer model Botfip relies on fundamentally analyzes global features of the OTS sequence, then contrasts it with Funcimg features through a multimodal alignment approach. Hence, Botfip essentially uses global information from Funcimg for OTS modeling, resulting in OTS sequences and operation tree structures that are not overly complex. In contrast, the GP and RL methods essentially iterate through trial and error using datasets to enhance fit. Under such circumstances, models are prone to generate more complex expressions for fitting, as illustrated by gplearn's result and many final outcomes of PYSR, where the solution with the smallest fitness often includes an extremely complex remainder, also indicative of overfitting.

A current issue with Botfip is its limited extrapolation capability, along with the absence of a viable error-correction mechanism. Unlike Langrange modeling, even minor errors in OTS modeling can lead to significant anomalies in the final results (such as a multiplication node in a binary node being incorrectly generated as a division node). This strictness leads to the model's insufficient extrapolation ability. Additionally, the lack of a correction mechanism also highlights Botfip's limitations in SR tasks. However, the high complexity of the OTS search space implies that Botfip requires more data for recognition.

\section{Conclusion}
Despite the current limitations of Botfip in extrapolation capabilities for SR problems, its essence lies in being a multimodal framework for Funcimg and OTS in scientific computing. It has the potential to evolve into a larger model through contrastive learning and the support of extensive data. With more computational power and data investment, Botfip's expressive capability and its performance in SR could be significantly enhanced. In the future, we also aim to further explore an automatic error-correction mechanism for OTS within the Botfip framework to better meet the needs of the SR field. Moreover, we plan to delve into the potential applications of the Botfip framework in other AI for Math and AI for Science problems.

\newpage
\bibliographystyle{named}
\bibliography{ijcai24}

\end{document}